\documentclass[referee]{cjaa}           

\usepackage{natbib,graphicx}                   
\input{epsf.sty}                        
\input{psfig.sty}                       

\begin{document}
\newcommand{\beq}{\begin{equation}}
\newcommand{\eeq}{\end{equation}}
\newcommand{\bu}{{\bf u}}
\newcommand{\bx}{{\bf x}}
\newcommand{\bk}{{\bf k}}
\newcommand{\bvv}{{\bf v}}

\title{Forecast for Epoch-of-Reionization as viewable by the PrimevAl
  Structure Telescope (PAST)}

   \volnopage{Vol.0 (200x) No.0, 000--000}      
   \setcounter{page}{1}          

   \author{Ue-Li Pen
      \inst{1,2}\mailto{}
   \and Xiang-Ping Wu
      \inst{2}
   \and Jeff Peterson
      \inst{3}
      }
   \offprints{U. Pen}                   

   \institute{
	     Canadian Institute for Theoretical Astrophysics,
             60 St. George St., Toronto, M5S 3H8, Canada \\
             \email{pen@cita.utoronto.ca}
        \and
National Astronomical Observatories, Chinese Academy of Sciences,
             Beijing 100012, China\\
        \and
             Carnegie Mellon University
             Pittsburgh, USA\\
          }

   \date{Received~~2004 month day; accepted~~2004~~month day}

   \abstract{
We present  sensitivity forecasts for the PAST 21cm reionization
observations and compare these with the measured sensitivity of a
prototype array.   
We discuss the likely epoch-of-reionization structures
and the redshifts that can be observed, and conclude that a 10,000
antenna PAST will be able to image ionized structures with its
1 million pixel map at $6<z<20$.  The  angular scales of the images to
be produced span from 5 arc minutes 
to 10 degrees.  The effect of potential foreground contaminants are
analyzed.  All known extra-terrestial foregrounds have power law
continuum spectra, and can be modeled and cleanly separated.  We present
results of observations using the prototype, 
which
verify the estimated sensitivities and provide data on astronomical foregrounds.
The same sample allows real-world measurement of man-made
interference, of meteor scatter propagation,  
and of ionospheric variation. 
These data are also used to verify our large angle map
making algorithms. 
   }

   \authorrunning{U. Pen, X. Wu \& J. Peterson }            
   \titlerunning{Forecast for PAST}  

   \maketitle

%
%
\section{Introduction}           
\label{sect:intro}


The frontier of optical observation of the universe is traced by the
furthest quasars ($z \sim 6.41:$\cite{2003AJ....126....1W}) and
galaxies \citep{2004A&A...416L..35P}.  Measurement of the absorption of
UV light from quasars allows the late ionization history to be traced.
This history has raised some apparent contradictions, which we will
discuss below.
Using VHF radio observations
of line emission it may be possible to directly examine ionization at
much higher redshifts-- perhaps to redshift 20.  With the PrimevAl
Structure Telescope (PAST) it should be possible to directly study the
Epoch of Reionization, mapping high redshift ionized structures in
three dimensions.

A definitive measurement of the epoch of reionizaiton (EOR) has many
implications.  It breaks 
some degeneracies in the CMB determinations of cosmological parameters
by providing a direct measure of optical depth.  By studying
ionization one learns how the dark ages ended, and  about the
energetics of the first objects. In addition, such observations
provide a very large number of resolved objects with direct redshift
measurements.  These can serve as source screens of beacons, allowing
mapping of density concentrations in the high redshift universe using
gravitational lensing \citep{2004NewA....9..417P}.

\subsection{History}

We start with a brief review of the related cosmic history.  After
recombination at $z=1100$, the universe enters a period popularly
called the 'dark ages'.  The CMB is a blackbody at $T=2.7(1+z)$K with
very little structure.  The neutral baryonic component expands
adiabatically with a kinetic temperature colder than the CMB by
$[(1+z)/1100]^{5/3}$.  The spin temperature of the 21cm transition is
initially in thermal equilibrium with the CMB, and decoupled from the
baryon kinetic temperature.

Around $z\sim 100$, the first non-linear objects form.  The gas cannot
cool radiatively, and acquires the same virial temperature as the dark
matter, well above the CMB temperature.  For the first such
'minihalos', the characteristic 
temperature is hundreds of degrees, at overdensities of several
hundred.  At such densities and temperature, the hydrogen nuclear spin
temperature 
couples to the kinetic temperature of the
gas \citep{1997ApJ...475..429M}.  By redshift 20, a major fraction of
matter is  in virialized gravitationally bound halos,
which, being warmer than the CMB, emit on the 21cm line. The minihalos are
optically thin, and we see these clouds spread out in redshift. So,
for observations  
that do not resolve the clouds, the 21 cm emission becomes an
effective continuum, with 
slight intensity variations that track the average density. In this
way the warm gas 
is observable via a brightness increment above the CMB.
The typical surface brightness is 23 mK\citep{2003MNRAS.341...81I}. 
  
The universe is ionized today, but the precise timing and
history of reionization is not known.  WMAP has measured the
temperature polarization cross correlation, which constrains the
reionization redshift $9<z_r<20$.  This constraint appears robust, but
at odds with quasar spectra which show the Gunn-Peterson effect
indicating a late reionization $z_r \sim
7$\citep{2003AJ....126....1W,2004astro.ph..1188W}.  CMB polarization
arises from Thomson scattering of CMB photons off free electrons,
which translates CMB quadrupole fluctuations into a polarization
signal.  Since the CMB fluctuation amplitude is known, the observed
polarization directly maps into a Thomson optical depth $\tau=0.17\pm
0.02$\citep{2003ApJS..148..161K}.  One needs to add all electrons up
to redshift $z\sim 15$ to accumulate this optical depth.

Quasars from SDSS shows Ly$\alpha$ completely absorbed
\citep{2003AJ....126....1W} at $6<z<6.3$.  Using the also fully
absorbed Ly$\beta$ line, they constrain the optical depth of
$\tau($Ly$\alpha)>22$.  For a neutral universe, $\tau_{\rm
GP}=3.42\times 10^5[(1+z)/7.08]^{3/2}$.  This translates into a
neutral fraction greater than 0.006\%.  If both measurements are
correct, the appearance of a neutral fraction above z=6, seen via
quasars, and the large ionized fraction implied by WMAP, then either
reionization proceeded very gradually or there was more than one
reionization\citep{2003AJ....126....1W,2004astro.ph..1188W}.  We
discuss these possibilities below.

Reionization has two potentially observable effects on the 21cm
brightness.  It changes the global sky spectrum
\citep{1999A&A...345..380S}, and also generates spatial structure
\citep{2004MNRAS.347..187F,2004NewA....9..417P}. With PAST we will
search for spatial structure.

\subsection{Cosmic Reionization Conundrum}

If the source of ionizing radiation is stars or quasars, the transition must
occur causally with localized sources. This presents a challenge: is it
possible to ionize the universe homogeneously, and therefore gradually, 
through an opaque
medium?  Studies of quasar spectra constrain the Ly$_\alpha$
optical depth, while it is the UV continuum optical depth that
controls ionization.  However, the equivalent width of the 
Ly$_\alpha$ line is comparable to
that in the continuum. For example, over the
optically thick redshift range $6.1<z<6.3$ observed by SDSS, the
total equivalent width on Ly$_\alpha$ is $ \tau($Ly$\alpha) \Delta z/z
\sim \tau_{\rm continuum}> 1$.  This suggests that it is
difficult for continuum photons to enter the dark regions in the
quasar spectra.  
Only a region containing a continuum source can
maintain a high ionization fraction.  
The
ionizing sources are localized.

We think of re-ionization occuring through many Stromgren
spheres, which grow and overlap.  Each Stromgren sphere is a region of
optically thin ionized material surrounding an ionizing source.  
UV continuum radiation propagates freely inside the sphere, but
cannot penetrate the surrounding material. The
transition from ionized and neutral is very sharp, unless the
radiation is very hard. These Stromgren spheres grow with time.  At
some point, they start overlapping, which marks the completion of
reionization.  This completion has some resemblances to the nucleation
of a first order phase transition.  When these spheres overlap, the
neutral fraction drops very suddenly.  In a single Stromgren sphere,
each ion sees just the radiation from its ionizing source.  Once the
spheres overlap, each atom sees all the ionizing sources all the way
to the Hubble radius.  Just as in Olber's paradox, the brightness is
now dominated by the far away sources.  The flux increases as the
ratio of the mean separation between ionizing spheres to the Hubble
radius.  This separation is most popularly modeled as the separation
between the most massive minihalos, which is several comoving
Megaparsecs.  At any rate, the separation would have to be
significantly smaller than the $\Delta z=0.2$ width of the SDSS
absorption trough.  The Hubble radius is two orders of magnitude
larger, so the neutral fraction in the ionized regions should drop by
two orders of magnitude at the end of the epoch of reionization,
thereby reducing the optical depth on the continuum to less than 1\%.  At
this point, the optical depth on the Ly$\alpha$ line should be
significantly lower than unity, at quite apparent odds with the SDSS
quasar data.

Other arguments based on the expansion speed of the Stromgren spheres
around the known quasars at $z>6.1$ also suggest a mostly neutral IGM at
these redshifts \citep{2004astro.ph..1188W}.

We are faced with a paradox that the SDSS quasar spectra imply an
ionization redshift close to 6, while WMAP implies larger than 10.
There are a couple of possible resolutions from this paradox.  One
possibility is that the Stromgren spheres are comparable in size to
the Hubble radius.  Another is that reionization occurred twice
\citep{2003ApJ...591...12C}.  In such a scenario, the first
reionization would occur at 
high redshift, say 20, possibly as a result of molecular hydrogen
cooling.  The ionizing sources turn off again, and
the hydrogen recombines.  It then reionizes again near redshift of 6,
perhaps as a result of instabilities arising from atomic hydrogen line
cooling.  Other more exotic possibilities include ionization from hard
X-ray or $\gamma$ rays, which would generate fuzzy boundaries in the
Stromgren spheres.  Uniform radiation from a decaying dark energy or
decaying dark matter would also not be subject to these constraints.

\subsection{An Opportunity}

Both theoretical analysis and the data indicate that the
ionization history may have been very complex.
A direct mapping of the neutral gas would provide the data
to the re-ionization debate.  Such data would help answer questions
about the epochs 
of reionization,
the physical scales of the process, and simply provide a new probe
into the high redshift 
universe.  In this paper we will outline the PAST strategy to achieve
this goal.  We note that the characteristic angular scales (5'-20'),
surface brightness (23mK), flux level (0.1 mJy), and spectral
resolution (300 khz) are all straightforward in modern day radio
astronomy.  Several existing arrays would have  enough
sensitivity to image reionization.  The only obstacle is the serious
man-made radio interference over the broad VHF band, which includes TV
and FM radio.

In this paper we will address questions of sensitivity, foreground
confusion, and data processing challenges.  We will describe the
prototyping campaign in section \ref{sec:prototype}.  In section
\ref{sec:strategy} we describe the observing strategy to minimize the
impact of foreground sources.  The foregrounds have spatial but not
frequency structure, while the ionization signals have both.  We
discuss below how to exploit this fact to achieve the required
sensitivities.  We have designed the PAST array to detect and map the
ionization signals.  PAST will use high gain phased arrays fixed
pointed at the NCP.

\section{Prototype and Site Test}
\label{sec:prototype}
A series of site tests were performed over the period of August
2003-April 2004  at three sites in the XinJiang Autonomous Region, one site in
Qinghai Province and at the South Pole.  All sites were visited and
tested for radio frequency interference (RFI).  
Each site has advantages, but at this stage of the project, accessibility is 
very important. We chose Ulastai in western China, at 90$^o$ E, 43$^o$ N,
and an altitude of 2600m for the current round of prototype testing.  
The site is surrounded by mountains over 4000m on all sides, which act
as an excellent ground shield against 
RFI. The nearest cities are Korla and Urumuqi, each about 200 km away.
The valley itself is served by a major railroad, and by roads that are
open all year.   
The railroad uses a narrowband FM communication channel at 137.44 MHz.
Electrical power mains provide AC power 
on-site. 
Cell phone towers serve the area, and service is good at the site, but the
cell-nodes and handsets but do not interfere at
the frequencies of interest.

\subsection{Prototype Setup}

The most extensive prototype testing was performed over the period March 1-5,
2004.  Four phased arrays (called pods) of 7 log periodic antennae
were set up on an 
approximate east-west baseline of 1.1km.  Two pods were near the
processing facility, and two more were connected by an intensity
modulation analog fiber 
optic link.  Logistic and technical support was provided by the Urumqi
radio observatory.

We have built and tested several antenna designs, including a small
loop antenna, a three octave log periodic dipole array, and a square
log spiral antenna.  For the test described here we used a two octave
log periodic dipole array.  The elements are horizontally
polarized. The longest element length is 1m measured from the central
transmission line to the edge, the self-similar parameter is
$\sigma=0.26$ and $\tau=0.76$.  The number of elements (9) was chosen
to allow high gain operations from 80-300 MHz. Numerical simulation of
the design, placed over a ground plane, indicates about 10 dbi gain.
Below 80 MHz the antenna has no resonant active region and behaves as
a short dipole close to the ground.  From 80 to 200 MHz the antenna
matches the 50 ohm feedline with SWR below 2:1. At our lowest
operating frequency, 50 MHz, the antenna presents impedance 2 - j 200
ohms, an inefficient match to the 50 ohm feedline. However we intend
to use amplifiers with noise temperature below 40 K, and the sky
brightness temperature at 50 MHz is about 4000 K so even with low
coupling efficiency sky noise will exceed the amplifier noise
contribution. 

\begin{figure}
   \vspace{2mm}
   \begin{center}
   \hspace{3mm}\psfig{figure=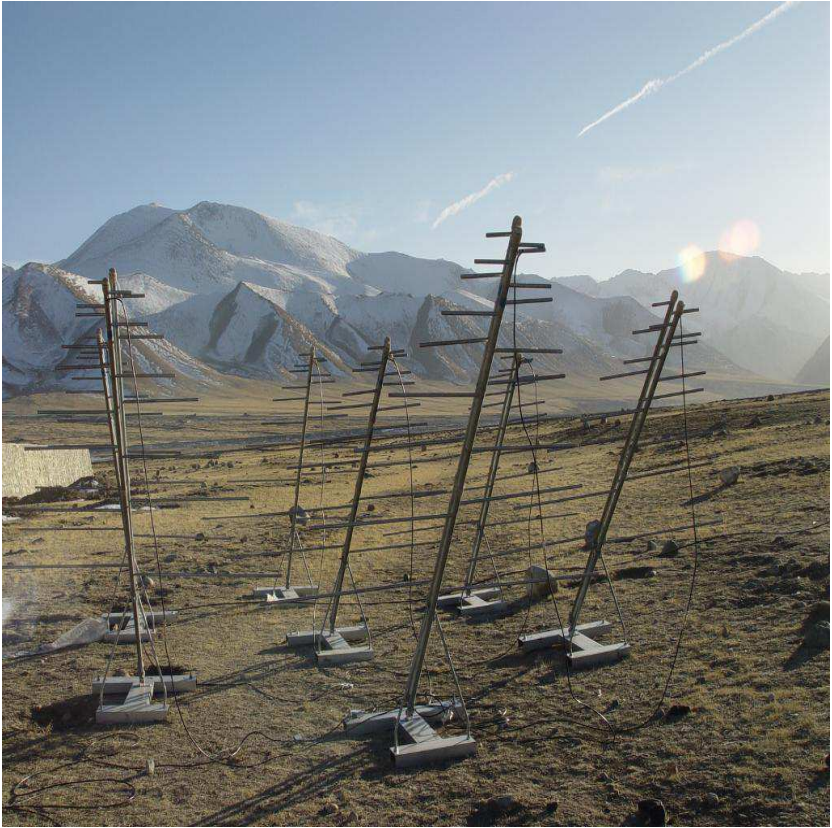,width=110mm,height=110mm,angle=0.0}
   \caption{Prototype pod layout.  
     The pod has an element separation
     between neighboring antennae of 2m when projected onto the NCP.
     The elements were summed with the appropriate delay lines to
     place the main lobe on the NCP.} 
   \label{fig:pod}
   \end{center}
\end{figure}

Each pod was a closed packed hexagon pointed at the north celestial
pole (NCP).  Figure
\ref{fig:pod} shows the layout.  The projected separation was 2m
for each side length.   The
phased array spacing is chosen to optimize performance near 100 MHz,
with some compromise of performance at the extremes of the range.  In
the current layout the effective area of the individual antennas does not
sum at low frequencies, and at high frequencies, array phasing
sidelobes appear. For our next design we plan to increase the high
frequency gain of the antenna to reduce the variation of effective
area with frequency.

The elements were summed in phase using coaxial cable delay lines and
commercial TV splitters, used in reverse as power combiners.  The
signal was bandpass filtered, amplified, 
and transmitted over the fiber link.  The data were digitized using a
commercial four channel 8-bit analog-digital converter board from
Acquisition Logic, which was sited in a dual Xeon 2.4 GHz server PC.

During operations, the westernmost pod did not work properly, however  14
hours of data were obtained on the remaining three baselines.
Observations started at 19:21 on March 3, 2004, Beijing time, and
ended at 11:42 on March 4, 2004. The results obtained from the data
are described in subsequent sections.

\section{Prototype Sensitivity}

Data from the prototype allows us to compare forecast
sensitivity with sensitivity measure using sources on the sky.  
In the maps we have created, the two
brightest sources within 10 degrees of the NCP are 3C061.1
at z=0.18 and 3C390.3 at z=0.56 These we detect at 34 and 38 sigmas
respectively.  Their flux at 75 MHz, interpolated from the 6C survey at
151 MHz and the 8C at 38 MHz are 59 and 69.4 Jy. At 75 MHz the array
has about 10.5 dbi gain. 
The effective area is then $A_{\rm eff}=G
\lambda^2/(4\pi)=14.1{\rm m}^2$.  Using our effective bandwidth of
$\Delta \nu=25$ MHz, this results in a net power of
$P_{3C061.1}=FA\Delta \nu/2=-128.9$ dbm through each pod for 3C061.
Slightly more  
uncertain is the noise of the system.  The galaxy has a surface
brightness on the pole of 1600K at 75 Mhz.  This contributes a noise
equivalent power $P_{\rm noise}=kT\Delta \nu=-95.3$ dBm.  Our summing elements
also had losses.  On a network analyzer, each summing element had an
insertion loss of 1.2 dB, and two summations were inserted.  The
amplifiers had estimated noise figures of 4 dB, and we budget an extra
3 dB for the remaining connector elements and impedance mismatches,
for a total noise figure of 
10dB.  This corresponds to another 1500K from the amplifier noise,
which doubles our effective noise budget.  Our total signal to noise
after 14 hours of integration is forecast to be $\eta \sqrt{2 t \Delta
\nu} P_{3C061.1}/P_{\rm noise}=66$ where we used $\eta=\sqrt{1/8}$ as
the correlator efficiency, and the factor of 2 in the square root
comes from the two correlation phases.  Our digital sampling board had
a net duty cycle of 1/4 due to the PCI bus transfer limitations, and
the correlation software had not been optimized on the single PC.
This estimate is close to the actual observed signal-to-noise.
The measured and forecast sensitivity match to within the uncertainty due to
loss of
coherence due to the ionosphere, 
and unknown impedance mismatches.

\section{Foreground Separation Strategy}
\label{sec:strategy}

The foregrounds exceed the reionization signal by at least a factor 1000.
To remove them we will need to use both spatial and spectral discrimination.
We list the foregrounds below along with our strategy for separation of the
signal form the foreground.

\subsection{EMI}

Since the band we will observe spans the FM broadcast, television broadcast and 
FM communications bands, man-made electromagnetic
interference (EMI) can be severe at many sites.  
We minimize EMI by using earth curvature and
mountain ranges.

\begin{figure}
   \vspace{2mm}
   \begin{center}
   \hspace{3mm}\psfig{figure=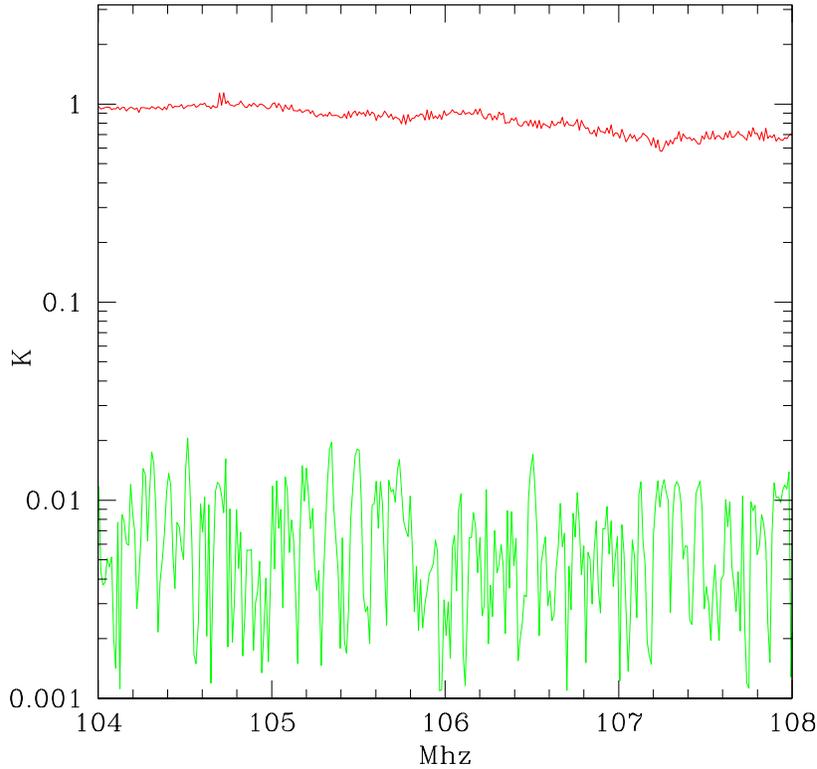,width=110mm,height=110mm,angle=0.0}
   \caption{Spectrum taken at Ulastai on Jan 2, 2004.  Shown is a
     short region in the FM band, which is usually the worst RFI
     regime.  The upper line is the noise after a 30 second
     integration, while the lower line shows the spectrum after an 8
     hour integration.  Only the frequency range 104-108 was plotted
     to make the structures visible. 
High intensity meteorite peaks have been
     cut. We see that no sources of interference are visible with an
     antenna temperature greater than 20mK.} 
   \label{fig:spectrum}
   \end{center}
\end{figure}

Figure \ref{fig:spectrum} shows the sky spectrum taken during a site testing
on Jan 2, 2004.  The commercial FM 88-108 MHz band is often considered
the worst for interference. We see no persistent sources down to 10
mK.   The only detectable source are intermittent transmissions from
the trains at 137 Mhz.
Residual EMI can be filtered, or mitigated
\citep{2000AJ....120.3351B}, but so far 
it has been sufficient to simply ignore the line at 137 MHz.  

\begin{figure*}
   \vspace{2mm}
   \begin{center}
   \hspace{3mm}\psfig{figure=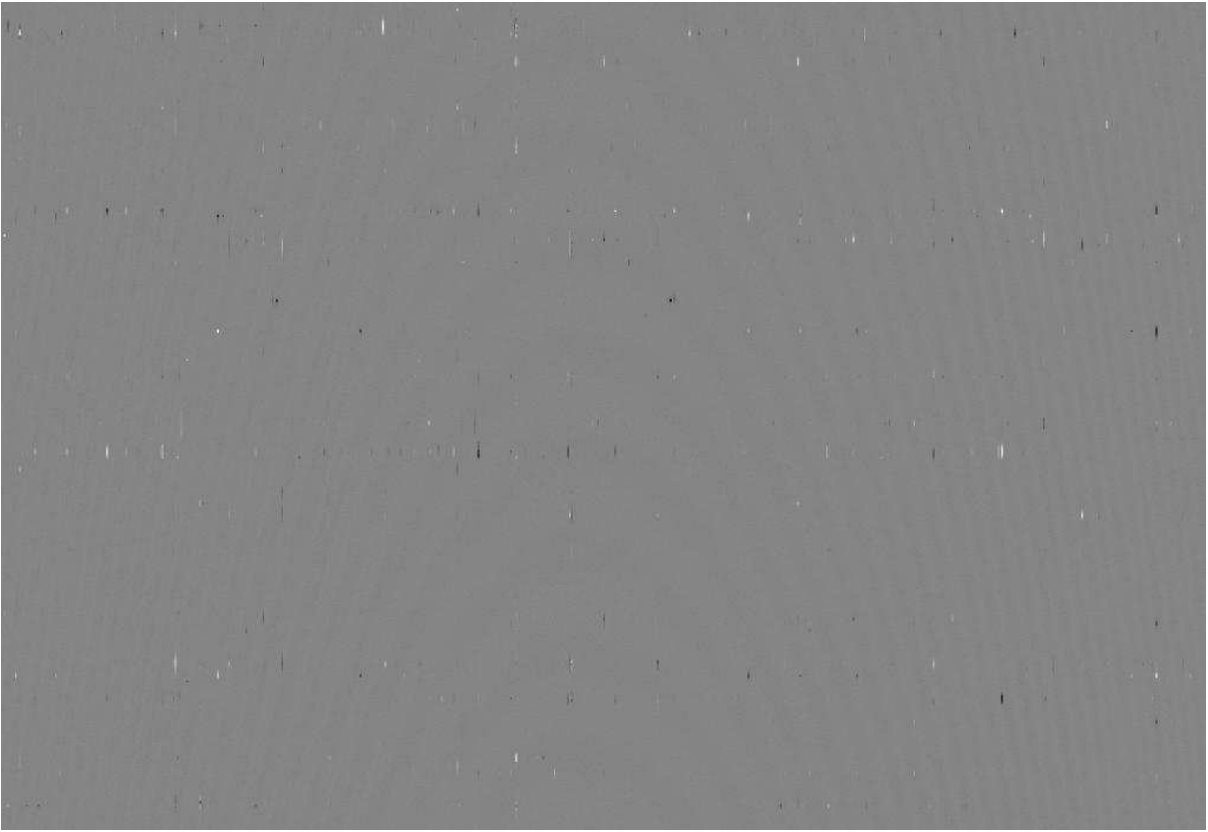,width=160mm,height=110mm,angle=0.0}
   \caption{RFI reflections from meteor trails as seen in the raw
     baseline data.  Time runs to the right, and a total of 
     4 hours are shown.  Frequency increases upward, and we see the
     range 75-82 MHz.  Only the real part of the correlation is shown,
     and the sign of the signal (black vs white) relates to the phase
     and direction of 
     the source.  Time resolution is 20 seconds.} 
   \label{fig:meteorite}
   \end{center}
\end{figure*}

When meteorites enter the upper atmosphere, they burn up and leave a
trail of ionized plasma.  The plasma frequency in this trail is
significantly higher than the general ionosphere, and can reflect
sources at ranges as far as 2200 km.  In the test data, we can clearly
see these meteor 
events.  Figure \ref{fig:meteorite}  shows the effect. 

Sporadically, when meteor trails allow propagation, RFI lines enter
the spectrum.   
Most of these events are unresolved in
time using our 20 second integrations, but there are a small number
which extend over more than 2 bins.  In these cases, the phases are
always changing 
across our bins, as would be expected for a meteorite traveling
more than our 10 arcminute fringe spacing in 20 seconds.

Eventually, these meteor trails can be tracked in real time in three
dimensions.  Multiple baselines allow a direct triangulation of the
height.  The 
velocity and direction of the trails can be reconstructed from the
data, if one samples more finely after a trigger.  It has been
proposed (Sasa Nedeljkevic, private communications) that these
meteor trails provide valuable information about the origin of the
particles.  PAST data will provide a large sample of data on the
distribution of extrasolar grain, 
including detail on
the trajectory and kinematics.

\subsection{Galaxy}

The sky brightness is dominated by synchrotron radiation from our Galaxy.
The surface brightness near the NCP is $T_{\rm
  gal}=280 (150 {\rm MHz}/\nu)^{2.5}$,
 
Radiometric fluctuations due to this emission are the 
primary source of noise in our observations: this brightness
determines the required integration time.   
Above the Galactic plane, the brightness is mostly
smooth, and interferometry allows one to integrate stably down from
the mean emission.  
The galactic synchrotron power spectrum above the galactic plane has
been measured by \citet{2002A&A...387...82G} based on the 2.4 Ghz
Parkes survey.  Its structure drops in $l^2 C_l \propto l^n$, with a
slope of about $n\sim -1.5$ to $l=1000$.  Extrapolating a power law to
the scales of interest, namely $l\sim 3000$, 
the galaxy has structure at the sub Kelvin level.
This still potentially leaves the galaxy 2 orders of magnitude brighter
in structure than the anticipated re-ionization signal.  Fortunately,
synchrotron 
emission has a mostly featureless frequency spectrum, which depends
on the energy distribution of relativistic cosmic ray electrons.
\citet{2003astro.ph.11514W} have discussed some of the issues.  We extend that discussion here.
We are looking for spectral structure with $\Delta
\nu/\nu<1\%$, over which range one can model out
the galaxy.  



\subsection{Point Sources}

This same strategy will also work against the third class of
foreground contaminants: point-like radio sources.  These
statistically dominate the power spectrum at $l>200$.  The bright
point sources can be subtracted, and their contribution to the total
power spectrum can be removed.  These will also serve as the primary
calibration source for the experiment.  One constructs a three
dimensional map, and adjusts the bright quasar continuum spectra as
the reference for a featureless power law.

Several other sources of noise are present, which can be addressed by
real-time data processing.  These include sferics (atmospheric
electric discharge like lightning) and meteor trails.  Meteor
trails generate higher density plasma which can reflect radio
transmitters that are normally not visible.  They last from small
fractions of a second to many minutes, and must be detected in the
data time stream and censored.  It helps to have instantaneous
location of point sources, which means covering two orthogonal axes
with the baselines.

\subsection{Ionospheric distortions}

The ionosphere can generate phase delays and scintillations.  The 6C
survey, which observed at 150 MHz, reported 5 degree phase changes on
km length baselines.  These can either be calibrated and corrected
using foreground sources, or ignored in a time average.

\begin{figure}
   \vspace{2mm}
   \begin{center}
   \hspace{3mm}\psfig{figure=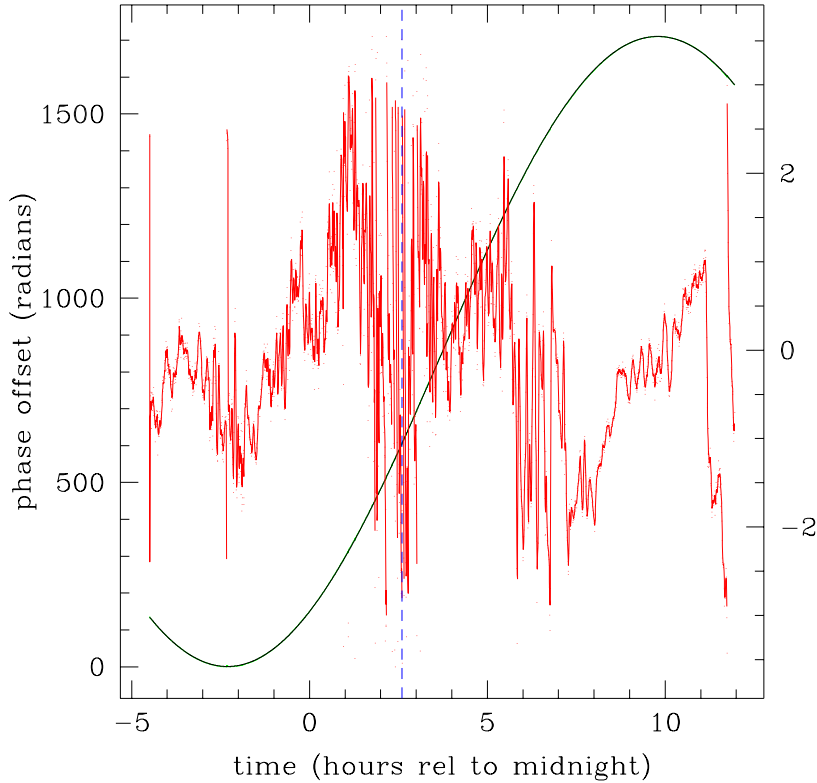,width=110mm,height=110mm,angle=0.0}
\caption{Phase delay of Cass A taken on a 1088m baseline on March
4, 2004. The horizontal axis is the time in hours.  The smooth
sinusoidal curve is the phase delay between the two baselines.
The residual points with the magnitude scale on the right is the phase
difference between the best fit sine wave.  The solid jagged line is a
4 point boxcar average, corresponding to a 100 second integration.  We
observe scintillations 
with characteristic periods of 20 minutes.  The dashed vertical line
is the meridian passage, when Cass A passes through a maximal air mass
of 4.4.} 
   \label{fig:cassa}
   \end{center}
\end{figure}

On the March 4 test run, we observed the phase of Cass A as it circles
the NCP. The result is shown in Figure \ref{fig:cassa}.
Usable data was obtained in the frequency range 62-88 MHz.
We observe typical scintillations with periods of 20 minutes, except
near the meridian crossing of Cass A.  At that point, Cass A passes
through the lowest point on the sky, at an elevation of 13 degrees,
corresponding to an air mass of 4.4.  This maximizes the impact of the
ionosphere, and also increases the projected phase fringe rate by the
air mass.

We find a characteristic wave length for the isoplanactic structure to be
20km.  This is significantly larger than the size of the array, and we
expect to be able to solve for the instantaneous phase delays for
bright sources using all pairs simultaneously.  The angular scale of
ionospheric structure would be 5 degrees, which is comparable to the
primary beam width.  If desired, one could build a model to compensate
for ionospheric variations.  There is also a degeneracy in our data
between spatial and temporal structure in the ionosphere.  It is in
principle possible that the ionosphere is varying in time, which could
contribute to the variations that we observe.  But the fact that the
variations increase with  airmass suggests that spatial variations
dominate.   The apparent velocity of the source
against the ionosphere is proportionate to airmass.

In practice, the ionospheric delays are small enough at the NCP that
they can be ignored without significantly affecting the image
quality.  Only for bright sources at large air mass, such as Cass A,
does one need to track and subtract. 

\begin{figure*}
   \vspace{2mm}
   \begin{center}
   \hspace{3mm}\psfig{figure=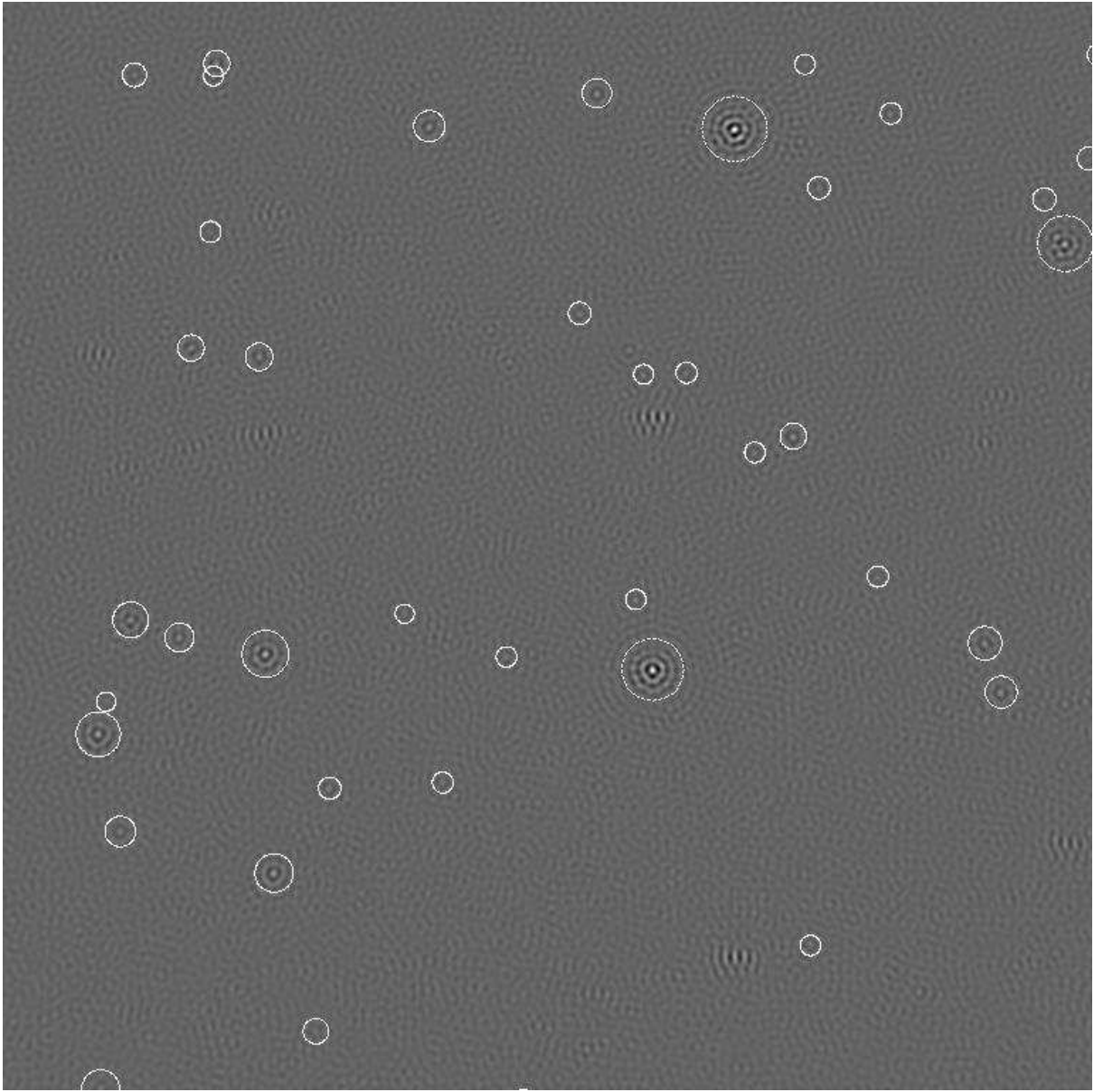,width=160mm,height=160mm,angle=0.0}
\caption{Image reconstructed using the 1088m east-west baseline.  The
displayed  field of view is 24 degrees.  Frequency range is 62-88
MHz. Overlaid on the gray scale
map from the PAST prototype are 8C sources with flux greater than 10
Jansky shown as circles.  The area
of each circle is proportionate to the 8C flux.  The adjoint map
making procedure brings objects into focus anywhere on the sky.} 
   \label{fig:test5}
   \end{center}
\end{figure*}

We show an image constructed from the test run data in Figure
\ref{fig:test5}.  The gray scale image used the data from one baseline
integrating over 14 hours.  The maps were constructed using the
optimal  techniques adopted from the CMB community, which is described
in further detail below.  In the small angle approximation, this is
analogous to the aperture synthesis dirty map.  The actual map making
procedure is described below, and includes several abberations from
small angle map making \citep{1992A&A...261..353C}.

\section{Design}
\label{sect:design}

We consider a general layout of $N_d$ antennae of effective aperture
$A_{\rm eff}$.  The effective aperture will be modeled relative to a
single dish in both beam size and geometric cross section.  For a
given angular resolution, we assume that the dishes are distributed as
a sparse subsampling of a single dish filled aperture reference
antenna with the corresponding sized primary beam.

Initially, the experiment will be sensitivity limited, so we consider
deep observations focused on the celestial pole.  Earth rotation
fills in the $u-v$ plane.  With $N_d$ elements, one obtains
$N_d(N_d-1)/2$ instantaneous baselines.  If one arranges them to cover
all separation lengths, earth rotation will result in all
angular configurations.  Specifically, if one achieves spacings that
are multiples of the effective aperture diameter, and fills all
baselines once, one obtains a resulting map which is spherical in the
$u-v$ plane containing $\pi N_d^2(N_d-1)^2/4$ pixels.  Expanding our
current prototype hexagonal geometry (Figure \ref{fig:pod}) to $n$
rings requires $1+3n(n+1)$ elements.  For $n=6$, we obtain a 127
element pod.  For an element spacing of 1.75m, the pod has a width of
20m. 

\begin{figure}
   \vspace{2mm}
   \begin{center}
   \hspace{3mm}\psfig{figure=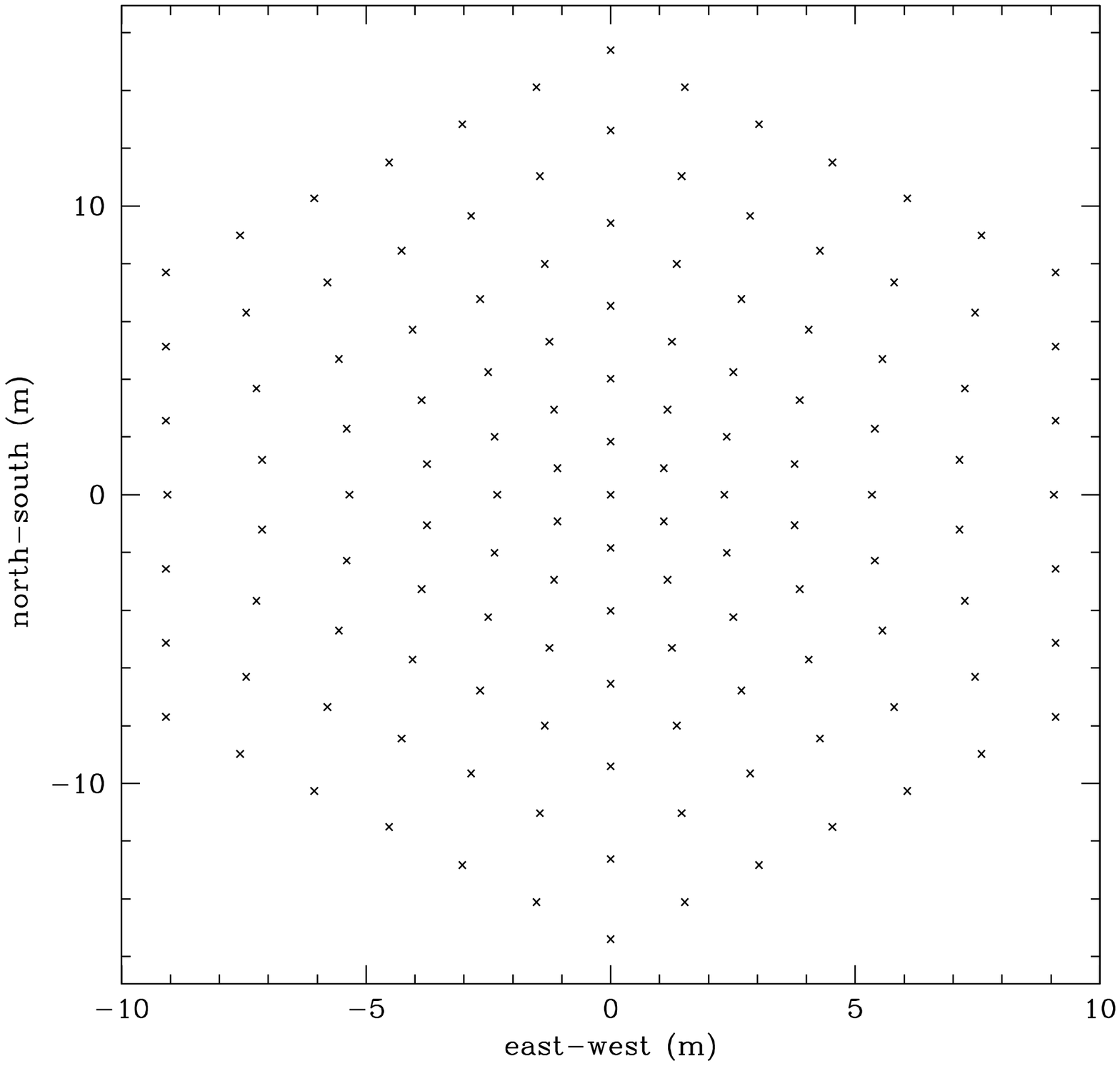,width=110mm,height=110mm,angle=0.0}
\caption{Rearranged elements with a 127 element pod to reduce the
strength of Bragg sidelobes.}
   \label{fig:podskew}
   \end{center}
\end{figure}

\begin{figure}
   \vspace{2mm}
   \begin{center}
   \hspace{3mm}\psfig{figure=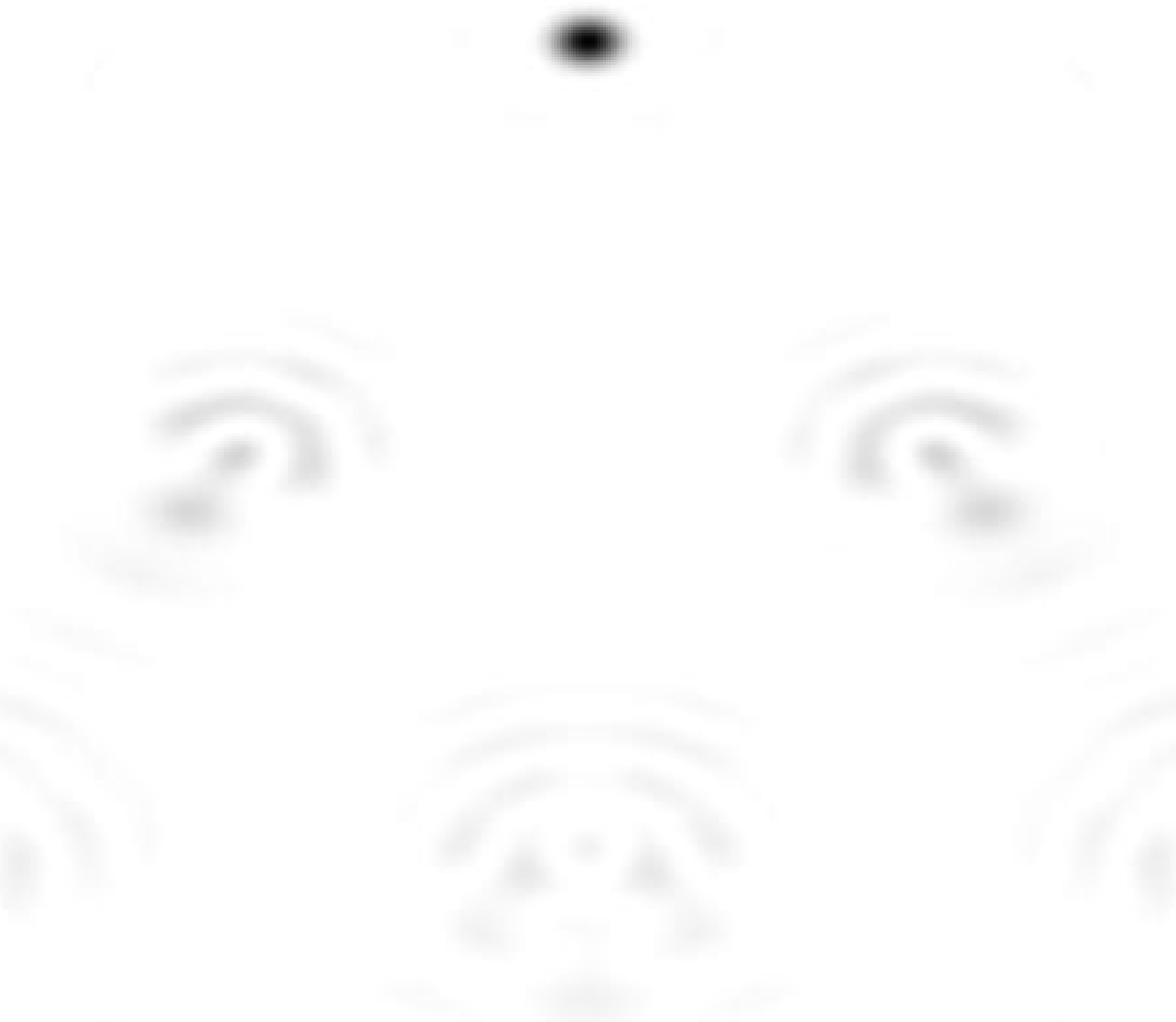,width=110mm,height=110mm,angle=0.0}
\caption{Polar projection of beam pattern for the array populated with isotropic
  receivers.  This shows the worst case side lobes, which occurs at
  the shortest wavelength, i.e. 200 Mhz. A spacing of 1.75m projected
  on the NCP was used.  The small residual side lobes are further
  suppressed by 20 dB 
  due to the primary beam of each log periodic.  The horizontal scale
  is 180 degrees, the vertical scale is 120 degrees.  The primary beam
  is actually round when viewed from the NCP.
}
   \label{fig:beampod}
   \end{center}
\end{figure}

A regularly spaced pod has Bragg reflection patterns at various
resonant angles across the sky.  These sidelobes can be suppressed by
a slight shuffling of spacings.  Figure \ref{fig:podskew} shows an
arrangement of the 127 elements in a pod which reduces the regular
first order resonances.  The resulting beam response in shown in
figure \ref{fig:beampod}.

The maximum size of the pods is determined by the need to achieve
redundant spectral imaging.  In order to overcome the bright
foregrounds, we will always be differencing maps at different
frequencies.  In order to do the differencing at fixed angular scale,
the baselines to be differenced must be of different physical scale.
The closer in frequency one applies the differencing, the less
susceptible one is to changes in spectral structure of the
foregrounds.  There is a lower limit, however.  The frequency
difference must be larger than the structures that we are searching
for.  This limits the largest possible frequency structures that one
can image.  We anticipate structures up to 40 Mpc in diameter
\citep{2004astro.ph..1188W}, which is 2\% in frequency.  In order for
the baselines to be distinct in the u-v maps, the pods should be no
larger than 30m for a 1.5km baseline.

With 40 elements, one obtains about 780 baselines.  One expects about
10 redundant baselines per non-redundant configuration if they
cover a 2km area and recover all the frequency information of a filled
aperture of 2km diameter.

\begin{figure}
   \vspace{2mm}
   \begin{center}
   \hspace{3mm}\psfig{figure=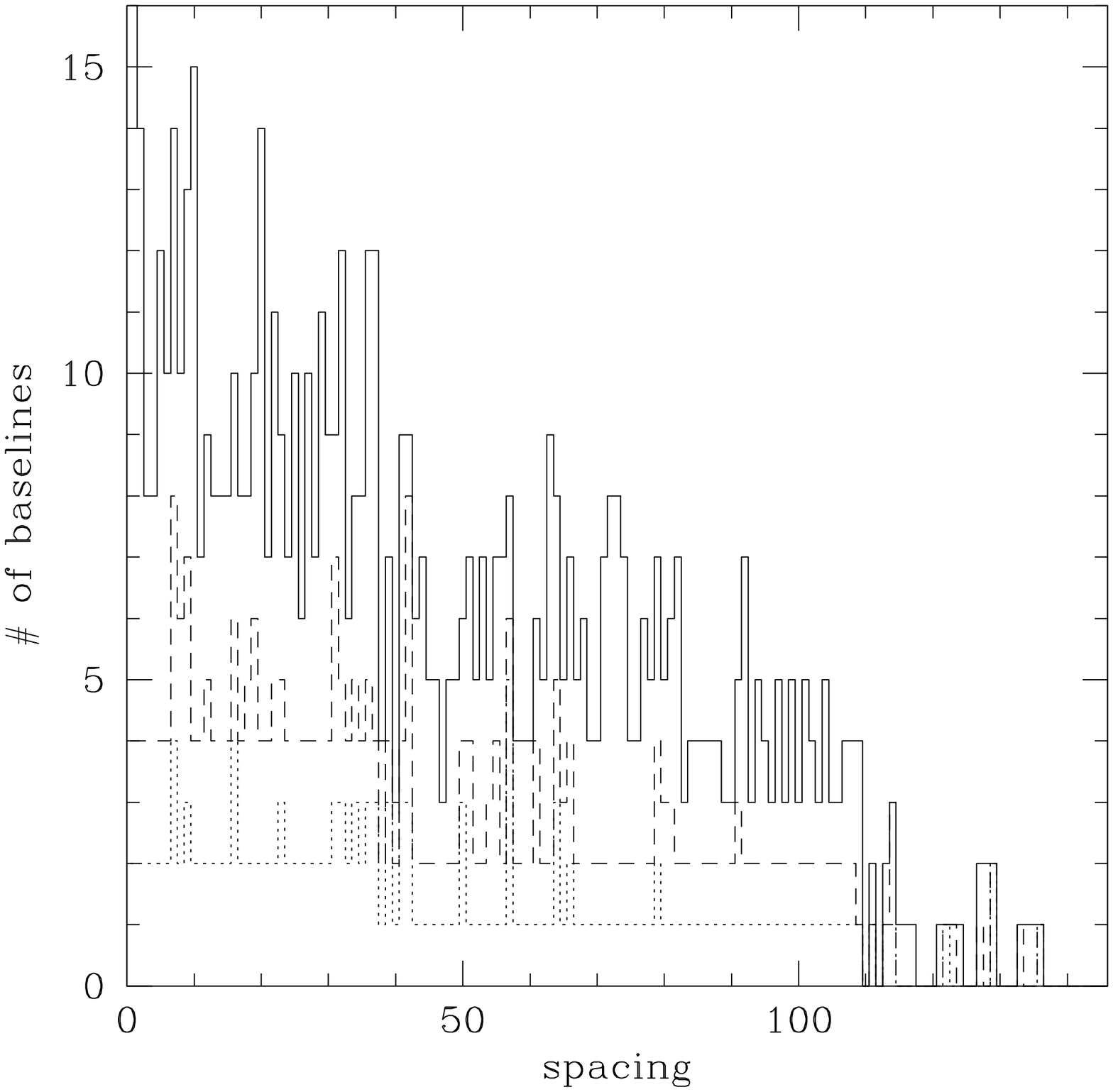,width=110mm,height=110mm,angle=0.0}
   \caption{
Baseline coverage with two sets of 10 pods offset by 73 pod units.}
   \label{fig:allbase}
   \end{center}
\end{figure}

Direct permutation layouts show that in one dimension, one can come to
within 20\% of a perfect non-redundant compact configuration.  This
means that 20\% of baselines are either redundant, or cover
non-contiguous parts of the $u-v$ plane.  Using two sets of 10 pods
over 67 pod diameters (which we call 'pod units') covers all baselines
configuration out to 36 pod units, including all correlations across
and within each size class.  We then duplicate this configuration
separated by 73 pod units, which covers all baselines up to 109 pod
units continuously, shown in Figure \ref{fig:allbase}.  The longest
baseline is 139 pod units.

\begin{figure}
   \vspace{2mm}
   \begin{center}
   \hspace{3mm}\psfig{figure=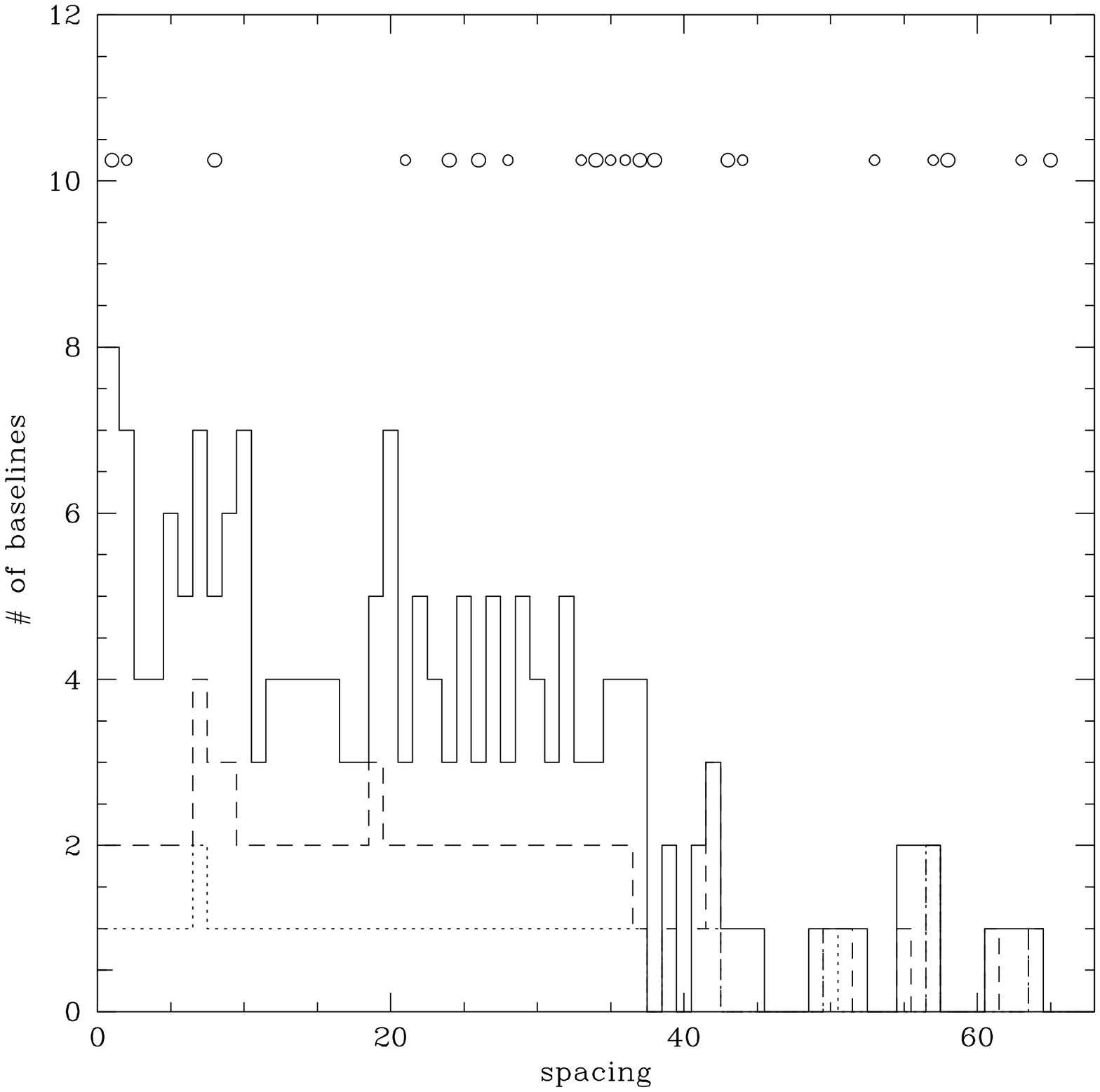,width=110mm,height=110mm,angle=0.0}
   \caption{Minimum redundancy configuration for two sets of 10 elements.  The
   combinatorial upper limit is 45 baselines covering every spacing
   from 1 to 45 exactly once.  This actual configuration covers all
   spacings up to 36 at least once, and is from an exhaustive search on all
   permutations.  The top row of circles denotes the location of pods,
   with large circles representing the initial pods and vice versa.
   The  histograms represent  the cumulative number of initial, secondary, and
   total baselines.}
   \label{fig:nonred}
   \end{center}
\end{figure}
Actual sensitivities depend on configuration, noise and angular scale.
An approximate non-redundant linear array of 10 elements from an
exhaustive search shown in figure \ref{fig:nonred}.  The actual beam
pattern of a rotation averaged non-redundant array is similar to a
filled aperture diameter corresponding to half the array size.  

To estimate sensitivities, we first estimate the sensitivity of a
single dish on a given length scale, and then scale by the dilution
factor of the array.  We will use 150 MHz as the central reference
frequency.  From the Haslam \citep{1981A&A...100..209H}
maps, we find the mean temperature in a 10 degree beam on the north
celestial pole to be 25K.  The spectral index near the NCP is 2.5 (see
below).  We expect a foreground noise temperature of $280 (150{\rm
MHz}/\nu)^{2.5}$K.  We budget a further 100K for receiver and system
noise.  The galaxy is the dominant noise source in the frequencies of
interest, 50-200 MHz.  The NCP is less than a factor of two warmer
than the coldest spots on the sky.

A reasonable estimate for the large scale strong features during
re-ionization might be 10 comoving Mpc, which is 5 arc minutes. Some
estimates are even larger than this value\citep{2003astro.ph.10338B}.
The observed quasars at z=6.28 and z=6.41 appear to have Stromgren
spheres with diameters of greater than 40 comoving
Mpc \citep{2004astro.ph..1188W}. 

Aiming at 10 $h^{-1}$ Mpc corresponds to
an aperture scale of 1.4 km.  One would expect features to have
comparable sizes radially.  The radial distance is 58 $h^{-1}$
km/s/Mpc (comoving) at $z=9$ for an $\Omega_0=0.3$ $\Lambda$-CDM
cosmology, so these 10 Mpc correspond to 300 kHz bandwidth.  The
sensitivity for a filled dish of diameter $d_f=1.4 (5'/\theta)$km is
\begin{equation}
\Delta T=2.5 \left(\frac{150 {\rm MHz}}{\nu}\right)^{2.5} \left(10 h^{-1}
{\rm Mpc}{L}\right)^{-1/2} \left( \frac{24h}{t}\right)^{1/2} {\rm mK}.
\end{equation}

The expected temperature change is 25 mK for an isolated Stromgren
sphere, so one achieves a S/N of order 10 in one day after a full
rotation of baselines.  The time increases as the array dilution
factor, so if one wishes a $1-\sigma$ map in four months, one can reduce
the collecting area from 1.5 km$^2$ to 1/100th that value.  If each
basic antenna has an area of 1.5m$^2$ (a gain of 7 dbi), this is
achieved with the 80 pod configuration of approximately 10,000
antennae. 

With a fixed number of antenna spread out over a 2km baseline, the
sensitivity increases rapidly for larger ionizing structures.  For the
array, we expect radiometric uncertainty
\begin{equation}
\Delta T=22 \left(\frac{150 {\rm MHz}}{\nu}\right)^{2.5} \left(20 h^{-1}
{\rm Mpc}{L}\right)^{-2.5} \left( \frac{24h}{t}\right)^{1/2} {\rm mK}.
\end{equation}
A week long observation should identify all ionized structures larger
than $20\frac{1+z}{10}$ Mpc at more than three sigma.  This
sensitivity is well matched to the inferred large ionization spheres
around SDSS quasars, which are twice as big and should be well resolved.

\begin{figure}
   \vspace{2mm}
   \begin{center}
   \hspace{3mm}\psfig{figure=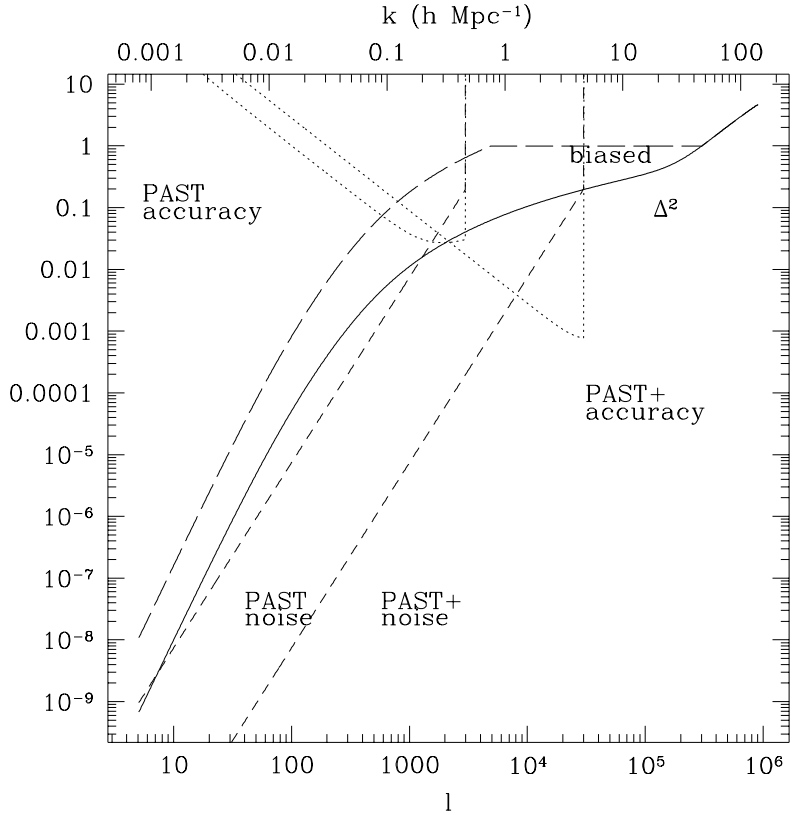,width=110mm,height=110mm,angle=0.0}
   \caption{
     Matter power spectrum measurement.
Solid line: the matter power spectrum at $z=9$.  One expects the 21cm
emission to trace the total mass.  Long-dashed line: power spectrum of
patchy re-ionization with a bias $b=4$.
Short-dashed lines: noise level expected
for PAST and second generation PAST+.  Dotted line: fractional
accuracy of power spectrum measurement
per logarithmic $l$ bin.
}
   \label{fig:power}
   \end{center}
\end{figure}

One can also aim for a statistical detection of the power spectrum.
If reionization occurs over the redshift range of $\Delta z \sim 2$,
we have 100 Stromgren spheres of $10 h^{-1}$ Mpc diameter radially.
If we use a pod beam of $10^o$, we expect about a million three
dimensional resolution cells of 10$h^{-1}$ Mpc.  One can measure the average
RMS variation between cells 30 times more sensitively than the actual
value of each cell.  With a total collection area of 15000 m$^2$, one
can measure a 10mK RMS power spectrum to 1000 $\sigma$ in 100 days.
This is quantified in Figure \ref{fig:power}.

To summarize:  50 pods are needed to cover the $u-v$ plane
completely.  A two dimensional structure is important to localize
emission.  At the same time, one needs many baselines with similar
length to be parallel to each other, so one can do simultaneous beam
frequency correction.  A T-shaped array with 50 elements on each axis
would satisfy such requirements. However, and array with fewer
elements may suffice if foreground removal can be accomplished using
incomplete uv coverage.

\section{Foregrounds}
\label{sect:foregrounds}

\subsection{Redundancy}

Each pod is a large array of phased antennae.  The maps are made from
baselines, and the sensitive measurements rely on a careful comparison
of spectral and spatial information.  This comparison will necessarily
require comparing differing baselines.  For example, a 1 km baseline at
150 MHz needs to be compared to a 1.1km baseline at 136 MHz to achieve the
same spatial scale.  While the baselines can be matched, the primary beams
may still differ.  This can be corrected for in the processing.  To
compare two maps with different beams, one convolves the first map
with the second map's beam and vice versa, so they have a common beam.
One strategy to cross check for beam systematics is to use two sets of
pods, possibly scaled by a small geometric factor.  This also makes a
natural deployment sequence, and we will discuss two generations of
pods, either one of which can make a complete image of the sky.
Individual elements within a pod may fail.
It is thus clearly important to be able to map the primary beam to
high accuracy.  Fortunately, the beams can be mutually calibrated.
Each baseline corresponds to factor of two range in angular scale over
the frequency band, which allows the construction of a mostly complete
u-v map from one baseline alone, as was done for Figure \ref{fig:test5}
using the prototype test data.  The grid of point sources as shown
in figure \ref{fig:6c} is mapped by each baseline.  The relative brightness
of the point sources measures the shape of the primary beam of the
product of the pods.  This allows one to solve for the primary beams,
given the redundancy in the number of baselines.

For example, if one element in one pod failed, one would find that all
baselines which involve this element would have a changed primary beam.
The primary beam is in turn a Fourier transform of the aperture, so one can
directly identify which element has failed.

\subsection{Radio Sources}

\begin{figure}
   \vspace{2mm}
   \begin{center}
   \hspace{3mm}\psfig{figure=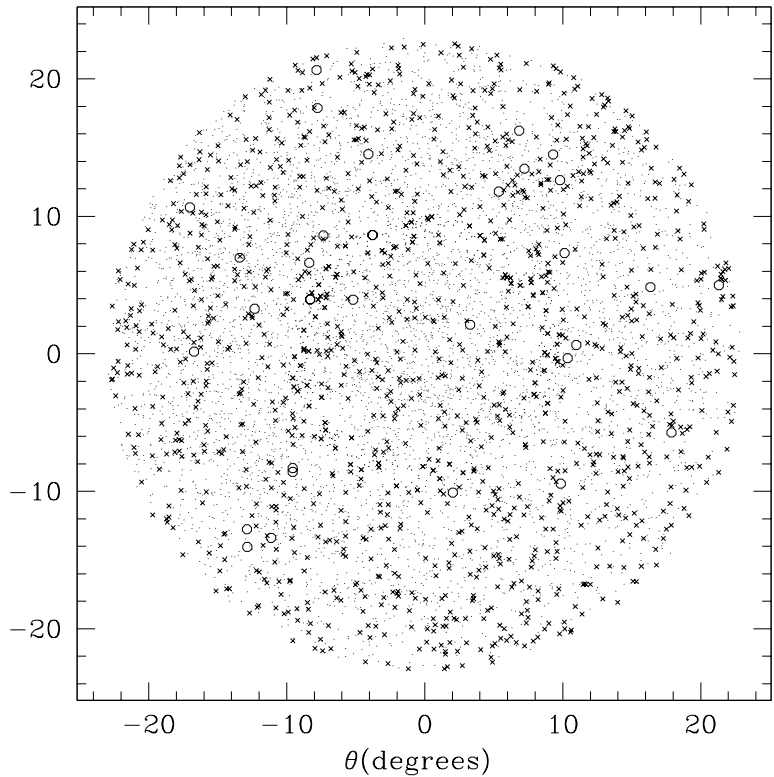,width=110mm,height=110mm,angle=0.0}
   \caption{ Distribution of 6C sources at 151 MHz around the NCP. Circles are
   fluxes greater 
   than 10 Jy, crosses between 1 and 10 Jy, and dots less than 1 Jy.
   The brightest source is 32 Jy.}
   \label{fig:6c}
   \end{center}
\end{figure}

The 6C survey of radio sources is based on an analogous concept of a
linear array of antennae at 151 MHz, which is in the middle of the
range of interest for PAST.  We can use it as a good guide to the
expected level of radio contamination at our frequencies of interest.
Figure \ref{fig:6c} shows the location of radio sources near the north
celestial pole.  The brightest source is 42 Jansky, and the source
counts are consistent with $d\log N/d\log S \sim -2.5$.  The total
flux is dominated by the collective faint sources.  The cutoff in this
integral is not well known.  At 150 MHz, the conversion from flux to
surface brightness is 1 Jy/sq degree=3.7K.  For a primary beam of 10
degrees, which corresponds to the central part of the map in Figure
\ref{fig:6c}, the  
total brightness contribution from resolved point sources in the 6C
survey would be 13K.  This is clearly a lower bound on the total
unresolved extragalactic background, which will contribute to Poisson
noise.  Extrapolating deep surveys to 6mJy at 610 MHz
\citep{1977IAUS...74...39W} results in a 48K background from point
sources.  Those counts appear convergent, and extrapolating to fainter
fluxes only adds 3K to the background.  The cosmic microwave
background adds another 2.7K.

At the confusion limit, we resolve sources down to 5', which should
resolve all sources down to 50 mJy.  While the total flux is dominated
by faint sources, the fluctuations are dominated by bright ones.
The residual fluctuations on 5 arcminute synthesized beams would then
be of order 25 K, which is expected to be larger than the fluctuation in
galactic structure.  The same procedure to remove the galaxy should
also apply to radio sources.  One could also add longer baselines to
further deconfuse the point source background.

%

\begin{figure}
   \vspace{2mm}
   \begin{center}
   \hspace{3mm}\psfig{figure=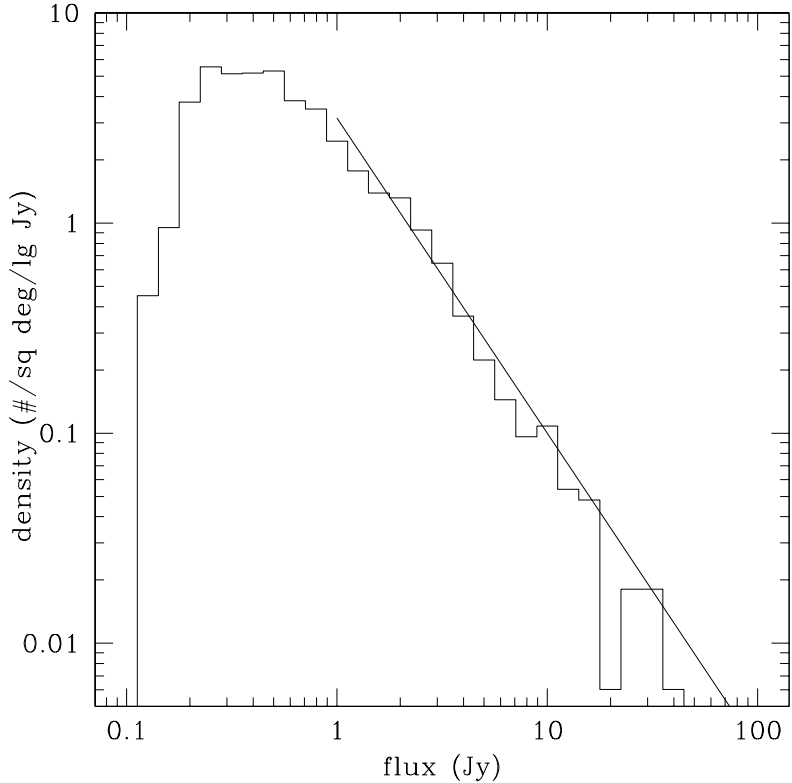,width=110mm,height=110mm,angle=0.0}
   \caption{Flux counts of 6C sources at 151 MHz.  The line segment represents
   $N(>S) \propto S^{-1.5}$.  The total flux is dominated by the
   unresolved faint end, while the fluctuations are dominated by the
   brightest.}
   \label{fig:counts}
   \end{center}
\end{figure}

\subsection{Galactic Emission}

The absolute temperature of the galaxy in the NCP region over the VHF
band has been well characterized over the last century.
\citet{1962MNRAS.124..459T} measured at 178 MHz, and find a surface
brightness of 175K at NCP.  They report a calibration error of 5\% and
a zero point error of less than 10K.  In the 408 Haslam maps, the same
point is reported at 25K.  Haslam reports a less than 10\% calibration
uncertainty, and a zero point error of 3K.  These numbers result in a
spectral index $T\propto \nu^\beta$ with $-2.7<\beta<-1.9$.  This is
significantly shallower than the values of $\beta=-3$ reported by WMAP
\citep{2003ApJS..148...97B} and Parkes \citep{2002A&A...387...82G}.
The low frequency synthesis of \citet{1987MNRAS.225..307L} report
$\beta=-2.5$ at the NCP.  We shall adopt that value for the frequency
range of PAST.  The total temperature budget scaled from 178 MHz is
then 280K at 150 MHz, which is dominated by the galaxy.

\section{Map Making}
\label{sec:maps}

Observing the North Celestial Pole is a form of drift scanning.  Drift
scan interferometry places unique requirements on the construction of
maps.  The CMB community has studied the general theory of map making
\citep{1997PhRvD..56.4514T}. With PAST we will have quite uniform uv
coverage, and we are searching for extended objects across the entire
field. The style of our observations and analysis is closer to CMB
observations than to typical radio observations, which may have
limited uv coverage, and anisotropic beams. For us, each baseline can
be thought of as generalized pixels with some noise and beam.  We will
simplify and apply CMB mapping theory to the interferometric
construction of EoR maps.

Each baseline can be considered a single pixel detector which drift
scans the sky.  The real and imaginary parts of the fringe can be used
to construct separate maps.  We can think of the visibilities for each
baseline and each frequency to be a set of generalized pixels.  We
denote them $\vec{v}$.  Their noise is uncorrelated, so the noise
covariance matrix ${\bf N}\equiv\langle v v^t\rangle$ is diagonal.
Each of these generalized pixels samples the sky through the
visibility beam.  We can think of this as a linear map $\vec{v}={\bf
P} \Delta$, where $\Delta$ is the set of temperatures at each point on
the sky.  The optimal map reconstruction is then $\tilde\Delta={\bf
P}^\dagger {\bf N}^{-1}\vec{v}$.  This map is the true map convolved
by the synthesized beam ${\bf B}\equiv{\bf P}^\dagger {\bf N}^{-1}{\bf P}$
so $\tilde{\Delta}={\bf B} \Delta$.  We see that the intrinsic
computational cost for the most general algorithm scales as the square of
the number of pixels.

The remainder of this section discusses how to implement this
procedure at a lower computational cost, and show some implementations
of algorithms such as CLEAN.  In the subsequent analysis,
we treat each pod as a uniformly filled dish.

\begin{figure}
\plotone{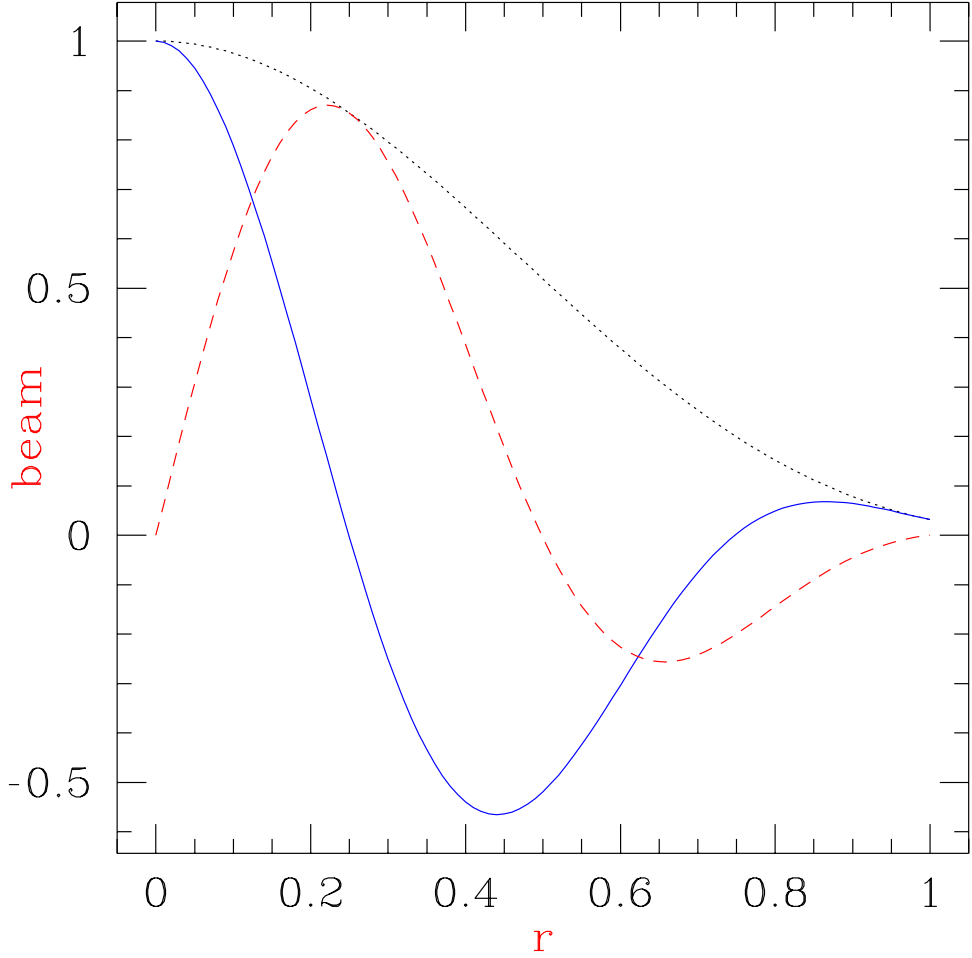}
\caption{Real space beam pattern of the correlation fringe of a
close packed baseline.  Scale is in units of primary beam FWHM.
The solid line is the beam from the real component of the correlation
fringe along the baseline vector.  The dashed line is the imaginary
component, and the dotted line the primary beam.  Perpendicular to the
baseline the window is determined by the primary beam.}
\label{fig:wreal}
\end{figure}

Let us first consider the real space window of a baseline.  We
approximate each pod as a circular filled aperture.  The real
component of the fringe with baseline along the 
$x$-axis results in a window 
\beq
W_R(x,y)=\cos(2 \pi n u_0 x)\frac{J_1(\sqrt{x^2+y^2}\pi
u_0)^2}{(x^2+y^2)\pi^2 u_0^2} 
\label{eqn:cos}
\eeq
in terms of the Bessel function $J_1$ with the imaginary component resulting in 
\beq
W_I(x,y)=\sin(2 \pi n u_0 x)\frac{J_1(\sqrt{x^2+y^2}\pi
u_0)^2}{(x^2+y^2)\pi^2 u_0^2} .
\label{eqn:sine}
\eeq
$n=1$ corresponds to the shortest spacing, and larger spacings
increase $n$ proportionately.  Figure \ref{fig:wreal} shows the of the
window function in real space.  We note that using the short
baselines, the full-width at half-max (FWHM) of the real component of
the beam is 0.41 of the FWHM of a single aperture with the same
diameter.  The FWHM is measured as the separation along the baseline
vector at which the real component beam drops from its peak halfway to
its global minimum.

We call the optimal weighted maps {\it raw
 map}, and proceed to fancier procedures including wiener
filtering and CLEAN.

\subsection{Raw Maps}

\begin{figure}
   \vspace{2mm}
   \begin{center}
   \hspace{3mm}\psfig{figure=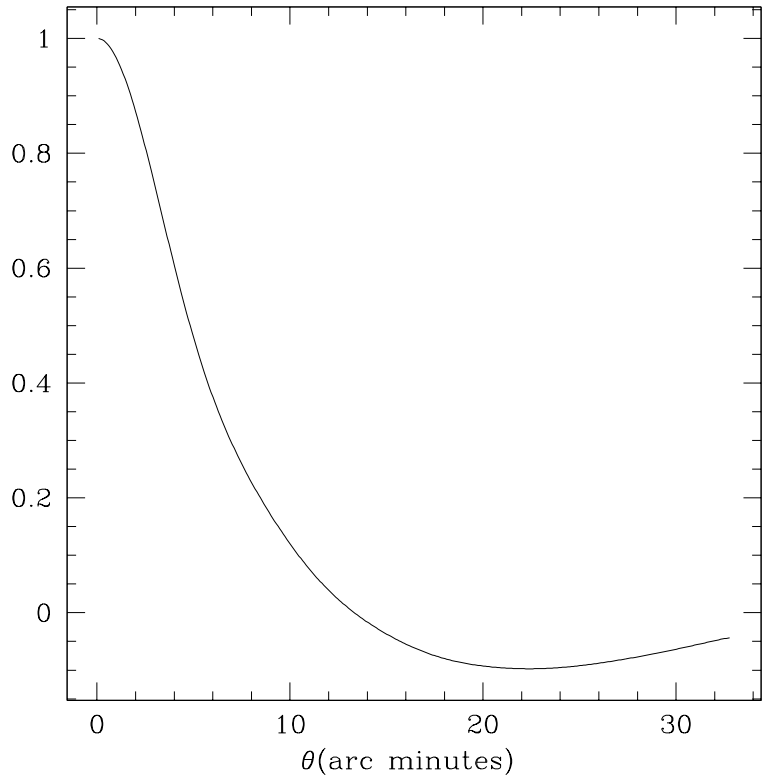,width=110mm,height=110mm,angle=0.0}
   \caption{Raw beam of PAST.  This raw beam is the autoconvolution of
   what is commonly called the resolution beam.}
   \label{fig:beam}
   \end{center}
\end{figure}

Let us normalize each base line to have unit noise per pixel.  Then
for a given baseline window (of which there are two per baseline, one
real and one imaginary) at an hour angle $\phi$, the visibility map is
given as $L_i(\phi) = 
\int d^3\bx W_i(\bx,\phi) S(\bx) + N$, where $S$ is the true sky 
map, $W_i$ is the $i-th$ window function normalized in such a way to
give unit noise variance in the map, and $N$ is white noise with unit
variance per pixel.  The raw map is then
\begin{equation}
\tilde{S}(\bx)=\sum_i\int W_i(\bx,\phi) L_i(\phi) d\phi
\end{equation}
We then define the {\it raw beam} to be
\beq
W_R(\bu)=\sum_i W_i(\bu)^2.
\label{eqn:natural}
\eeq
This raw beam transformed back to angular space is shown in Figure
\ref{fig:beam}. 
We will subtract the
mean signal from each scan since that contains ground, cross talk and
other stationary contaminants. This effectively slightly modifies the
window functions. 

\subsection{Co-linear Base Line Maps}

The fractional bandwidth of PAST is large.  Most sources are synchrotron
emitters with similar spectral index.  We can then treat the different
frequencies as different effective baseline lengths.  The fixed
baseline on earth measured in wavenumbers is a function of frequency,
and we can get a factor of two or more in scales from one baseline
alone.  The prototype map in Figure \ref{fig:test5} was made from a
single baseline.  Instanteneously one obtains a one dimensional map,
as a function of wavelength. Earth rotation maps this one baseline
onto all angles, resulting in a two dimensional map.  In the full
array, one will be able to simultaneously construct a set of co-linear
baselines at each frequency independently, resulting in an
instantaneously two dimensional map (spatial and frequency), which
earth rotation will fill in as the final full three dimensional map.

The baseline visibility at a given frequency sees the sky multiplied
by the primary beam, projected along the baseline, and Fourier
transformed.  The adjoint operation is the inverse Fourier transform,
projection onto the sky, and weighting by the beam.  The projection is
a scatter operation, which we can juxtaposition into a gather: each
pixel on the sky samples the Fourier transform of a baseline at each
position angle.  

\subsection{Cleaned Maps}

Densely sampled drift scanning differs fundamentally from image
reconstruction in traditional radio interferometry.  Typically, the
u-v plane is sparsely sampled, and images can never be fully
reconstructed even in principle.  Statistical techniques, such as
Bayesian priors or maximum entropy methods are employed to fill in the
missing information.

For fully sampled arrays like PAST, there are no gaps on the u-v
plane, and the synthesized beam is very ``clean'', i.e. does not
possess significant side lobes, as can be seen in figure
\ref{fig:beam}.  For non-overlapping point sources, the raw maps are
the likelihood function.  Peaks in the raw maps can then be identified
as point sources.  The signal-to-noise is easily known, and can be
read off from the diagonal elements of the noise variance matrix
$\langle \tilde \Delta \tilde 
\Delta^t \rangle = {\bf P^\dagger N^{-1} P}$, which is linearly proportial
to the diagonal elements of the beam matrix ${\bf B}$.

But we may still be concerned about point sources
that are ``noise'' in the quest for intrinsic 21cm fluctuations.  It
is desirable to be model point sources to as low a flux limit and as
high an accuracy as possible.

\subsection{Homogeneity of an east-west array}

It has long been known that a pure east-west array does not have most
of the issues that we have discussed above.  This is because fringe
patter of an E-W baseline is symmetric about a N-S line on the sky.  A
N-S baseline is inclined with respect to the line of site (except for
an array site at a Poles of the Earth). Fringes in the Northern half
of the fringe pattern are spaced closer together than those in the
Southern half.

Many of the
early radio astronomy arrays were layed out along an east-west
direction for this reason.  For a co-linear array, one can lay the
baselines into a u-v plane, and apply a fast Fourier transform.  The
real space map is then the celestial sphere projected down the NCP,
and the interferometric beam is homogeneous.  Our prototype pods were
11 degrees off from the perpendicular line to the NCP, and the non-flat
sky problems arose even 10 degrees from the NCP.

An east-west array has the benefit that the far side lobes of bright
sources will be easier to understand.  Also, being co-linear, the
anisotropies in the primary beam always have the same alignment
relative to the fringes.  Searching for small spectral structure
against bright foregrounds, this allows for a cleaner subtraction of
neighboring spectral maps.  As the earth rotates, bright sources can
cross the limb of the Earth, the ionosphere can change in absorption
or refraction, man-made EMI can change, and the instrumental offsets
can drift.  If all baselines lengths are observed at the same time in
the same position angle on the sky, these effects all cancel when
constructing difference maps at neighboring frequencies.

\section{Global All-Sky Spectra}

In addition to imaging the ionization structure, one could use PAST as
a tool to measure the absolute spectrum of the sky.  The global
reionization of the universe imprints a 20 mK feature on the global
diffuse spectrum.  An array of 100 elements each taking independent
spectra would achieve a sensitivity of 20 mK over a 1 MHz bandwidth at
150 MHz in less than two minutes.  If reionization completed rapidly,
as is generally believed, the small global signal can be measured with
PAST.

\section{Conclusions}

We have reviewed the current state of our understanding of
re-ionization, which points to a complex history with ionization
structure likely to exist on a range on length and time scales.  The
ionized regions are visible as holes in emission against the neutral
background, which can be imaged interferometrically.  

Estimates of the reionization scenarios show that the structures are
at frequencies, angular scales and flux levels that are 
accessible to existing radio astronomy facilities.  The only obstacle
is the serious level of radio interference on the VHF band, which
includes commercial TV and FM radio.

We have
presented estimates of the strength of the signal for currently best
guess models of the Epoch-of-Reionization scenarios.  The PAST
experiment should have sufficient sensitivity to image the larger
structures with diameters of $20 h^{-1}$ Mpc, and one can obtain 3-D
maps with high signal-to-noise over a period of weeks.  Smaller scales
can be probed statistically, in analogy with the current weak lensing
and CMB surveys.  The PAST maps will have up to 2.5 arc minute
resolution with a field of view of 5 degrees, and at least 10,000
resolution in redshift.

We have investigated the effects of various sources of noise and
foregrounds.  The dominant noise source is the galaxy, which makes the
remaining instrumental noise characteristics easy to control.  Other
sources, including man-made interference, meteor trail reflections,
extra-galactic radio sources, and others, have been studied in the
PAST protoype.  All celestial sources have featureless continuum
spectra, while the reionization sources have spectral structure.  All
prototype data indicates that the sensitivities should be
realistically achievable.
We have addressed several processing challenges, and demonstrated
efficient map construction on the curved sky.

\begin{acknowledgements}
This work was funded by the Natural Science Foundation of China (NSFC).
\end{acknowledgements}

\appendix                  

\bibliography{pastbib,penbib}
\bibliographystyle{mn2e}

\label{lastpage}

\end{document}